# ANGULAR DISTRIBUTIONS OF MASSIVE QUARKS AND LEPTONS CLOSE TO THRESHOLD†

S.J. Brodsky [b], A.H. Hoang [a,‡], J.H. Kühn [a,b] and T. Teubner [a,‡]

[a] Institut für Theoretische Teilchenphysik, Universität Karlsruhe,
D–76128 Karlsruhe, Germany

[b] Stanford Linear Accelerator Center
Stanford University, Stanford, CA 94309

**Abstract**

Predictions for the angular distribution of massive quarks and leptons are presented, including QCD and QED corrections. Recent results for the fermionic part of the two-loop corrections to the electromagnetic form factors are combined with the BLM scale fixing prescription. Two distinctly different scales arise as arguments of $\alpha_s(\mu^2)$ near threshold: the relative momentum of the quarks governing the soft gluon exchange responsible for the Coulomb potential, and a large momentum scale approximately equal to twice the quark mass for the corrections induced by transverse gluons. Numerical predictions for charmed, bottom, and top quarks are given. One obtains a direct determination of $\alpha_{\rm V}(Q^2)$, the coupling in the heavy quark potential, which can be compared with lattice gauge theory predictions. The corresponding QED results for $\tau$ pair production allow for a measurement of the magnetic moment of the $\tau$ and could be tested at a future $\tau$-charm factory.

*The complete postscript file of this preprint, including figures, is available via anonymous ftp at ttpux2.physik.uni-karlsruhe.de (129.13.102.139) as /ttp95-26/ttp95-26.ps or via www at http://ttpux2.physik.uni-karlsruhe.de/cgi-bin/preprints/ Report-no: TTP95-26.

†Work supported by BMFT 056 KA 93P.

‡e–mails: hoang@ttpux2.physik.uni-karlsruhe.de and tt@ttpux2.physik.uni-karlsruhe.de

## 1. Introduction

QCD predictions for the angular distributions of partons can be tested in a variety of ways. For massless partons, *i.e.* light quarks or gluons, the canonical approach is based on angular distributions of jets or related quantities, like the thrust axis. In fact, the observation of a distribution $\propto (1+\cos^2\theta)$ for two-jet-events has provided convincing evidence for the spin-1/2 nature of partons produced in $e^+e^-$ annihilation. Alternatively, for charmed or bottom quarks one may directly identify the heavy quark and heavy meson directions. This approach has been employed frequently in the analysis of the $b$ quark angular distributions at high energies. In this kinematical region gluon radiation affects the shape of the distribution, typically rendering the lowest order predictions more isotropic. Real and virtual corrections must be considered jointly, since only the sum leads to an infrared finite prediction.

In this paper a different kinematical region will be considered: heavy quark production close to threshold, where the cross section is dominated by a few exclusive channels. In this case one expects that the heavy mesons will essentially follow the heavy quark direction—a simple consequence of Newton's law of inertia. To employ the parton model in the case of hadron pair production may be somewhat surprising, since the structure of form factors for exclusive channels imposes stringent constraints on the angular distributions of the hadrons. However, as a consequence of the near degeneracy of the pseudoscalar and vector mesons $B$ and $B^*$, the sum of the various channels ($B\overline{B}$, $B\overline{B}^*$ + c.c., $B^*\overline{B}^*$) may easily combine to saturate the prediction for the angular distribution of the quarks. For definiteness, we will subsequently discuss the case of $b$ quarks. For certain kinematical regions our calculation can be applied for charm quarks and will be valid *a forteriori* for top quarks.

Close to threshold, in the limit $\beta = |\vec{q}|/\sqrt{\vec{q}^{\,2}+m_Q^2} = \sqrt{1-4m_Q^2/s} \to 0$, the cm angular distribution for $e^+e^- \to Q\,\overline{Q}$ is of course isotropic, a result of $S$-wave dominance. The small admixture of $P$-waves slightly above threshold provides a contribution $\propto \beta^2\cos^2\theta$ which will be studied in this paper. The tree graph prediction for the angular distribution is trivial. The leading $\mathcal{O}(\alpha_s)$ corrections have been evaluated in [1] for arbitrary quark mass and energy. However, unavoidably, the renormalization scale at which the strong coupling $\alpha_s(\mu^2)$ has to be evaluated in these formulae can only be fixed by a two-loop calculation. In the high energy region one may convincingly argue on physical grounds that the scale for the dominant correction terms should be of order $\sqrt{s}$. In the threshold region, however, both $\sqrt{s}$ and the quark momentum $|\vec{q}|$ are viable options for the energy scale. In the BLM procedure [2] the renormalization scale is fixed by resumming all terms proportional to the QCD beta-function into the running coupling. To carry out the BLM procedure to leading order, it is sufficient to identify the $n_f$ terms in the next to leading order coefficients. Recent results for the fermionic part of the two-loop corrections to the total rate [3], and in particular to the $\gamma^* \to Q\,\overline{Q}$ form factors allow for a solution of the scale-setting problem, at least in the threshold region. In fact, employing the BLM scale-setting prescription, one may even fix not only the scale but perhaps even accommodate the bulk of the two-loop corrections.

Real radiation is strongly damped in the threshold region, decreasing $\propto \beta^2$ relative to the tree contributions. Hence, dropping the terms of higher power in $\beta$, one may dispose of real radiation (even if multiplied by an infrared cutoff) and limit the discussion to virtual radiation; *i.e.* to the corrections to the form factors. Contributions from the instantaneous Coulomb potential and from hard photon exchange can be clearly distinguished, allowing for the resummation of the former. The contribution from the Coulomb part of the heavy quark rescattering yields a series in $(\pi\alpha_s/\beta)^n$. If one utilizes the $\alpha_{\rm V}$ scheme, defined for the heavy quark potential $V(Q^2) = -4\pi\, C_F\, \alpha_{\rm V}(Q^2)/Q^2$, then we find that the scale of $\alpha_{\rm V}$ is set to a value proportional to $s\,\beta^2$, the square of the relative momentum. Thus one can obtain a determination of $\alpha_{\rm V}(Q^2)$ from measurements of $e^+e^- \to Q\,\overline{Q}$ near threshold. In particular we shall show that the "anisotropy" $A(\beta^2)$ defined by

$$\frac{\mathrm{d}N}{\mathrm{d}\cos\theta} \propto 1 + A\,\cos^2\theta \qquad (1)$$

is sensitive to of $\alpha_{\rm V}(s\,\beta^2)$. Precise predictions for $\alpha_{\rm V}(Q^2)$ have been given in [4] using heavy quark lattice gauge theory with constraints from the $\Upsilon$ spectrum.

We also will discuss the application of the anisotropy formalism to the QED process $e^+e^- \to$

$\tau^+\tau^-$. We shall show that the anisotropy of the angular distribution provides a new way to measure the Pauli form factors of the leptons, the analog of the anomalous magnetic moment in the timelike region. In addition, the radiative corrections we discuss here change the normalization of the cross section near threshold.

## 2. Form factors, angular distributions, and infrared singularities

The amplitude for the creation of a massive fermion pair from a virtual photon is characterized by the Dirac ($F_1$) and Pauli ($F_2$) form factors:

$$\overline{u}\,\Lambda_\mu\, v = ieQ_f\,\overline{u}\,[\,\gamma_\mu\, F_1(q^2) + \frac{i}{2\,m}\,\sigma_{\mu\nu}\,q^\nu\, F_2(q^2)\,]\,v \tag{2}$$

where $\sigma_{\mu\nu} = \frac{i}{2}[\gamma_\mu, \gamma_\nu]$. The photon momentum flowing into the vertex is denoted by $q$, the fermion mass by $m$. The resulting angular distribution is conveniently expressed in terms of the electric and magnetic form factors $G_e$ and $G_m$ [5]:

$$\frac{\mathrm{d}\,\sigma(\mathrm{e^+e^-} \to \mathrm{f\bar{f}})}{\mathrm{d}\,\Omega} = \frac{\alpha^2\, Q_f^2\,\beta}{4\,s}\left[\frac{4\,m^2}{s}\,|G_e|^2\,\sin^2\theta + |G_m|^2\,(1+\cos^2\theta)\right] \tag{3}$$

with

$$G_e = F_1 + \frac{s}{4\,m^2}F_2\,, \qquad G_m = F_1 + F_2\,. \tag{4}$$

The anisotropy is thus given by

$$\begin{aligned} A &= \frac{|G_m|^2 - (1-\beta^2)|G_e|^2}{|G_m|^2 + (1-\beta^2)|G_e|^2} \\ &= \frac{\widetilde{A}}{1-\widetilde{A}}\,, \end{aligned} \tag{5}$$

where

$$\widetilde{A} = \frac{\beta^2}{2}\,\frac{|F_1|^2\,(1-\beta^2) - |F_2|^2}{|F_1+F_2|^2\,(1-\beta^2)}\,. \tag{6}$$

In Born approximation $A_{\mathrm{Born}} = \beta^2/(2-\beta^2)$. Note that for $F_2 = 0$, the anisotropy is identical to the Born prediction, independent of $F_1$. Thus the form

$$\frac{\widetilde{A}}{\widetilde{A}_{\mathrm{Born}}} - 1 = -\,2\,F_2(s)\left[1 + \mathcal{O}(\frac{\alpha}{\pi})\right], \qquad \widetilde{A}_{\mathrm{Born}} = \frac{\beta^2}{2} \tag{7}$$

isolates $F_2(s)$. This provides a way to experimentally determine the timelike Pauli form factor of the $\tau$ lepton. The QED prediction is

$$F_2(4\,m^2) = -\frac{\alpha}{2\,\pi} + \mathcal{O}\left(\left(\frac{\alpha}{\pi}\right)^2\right) \tag{8}$$

which is, up to the sign, equal to the familiar Schwinger result $F_2(0) = \alpha/2\pi$. Away from threshold the one-loop QED prediction is

$$F_2(\beta) = \frac{\alpha}{\pi}\left[\frac{1-\beta^2}{4\,\beta}\,\ell n\frac{1-\beta}{1+\beta}\right]. \tag{9}$$

We neglect this type of higher twist corrections in the following.

In Born approximation $F_1 = 1$ and $F_2 = 0$. The impact of one- and two-loop radiative corrections on the form factors and angular distributions will be discussed in the context of QED in the remainder of this section; the case of immediate interest, namely QCD, will be discussed in section 3. To demonstrate the line of reasoning, we first shall present the arguments for the leading order calculation in some detail.

In the QED calculation the order $\alpha$ correction to the Dirac form factor $\delta F_1$ in the timelike region exhibits an infrared singularity which can be regulated using a nonvanishing photon mass $\lambda$:

$$\delta F_1 = \delta F_1^{\rm fin} + \delta F_1^{\rm IR}\,\ell n\frac{s}{\lambda^2} \tag{10}$$

with

$$\delta F_1^{\rm fin} = \frac{\alpha\,\pi}{4\,\beta} - \frac{3}{2}\frac{\alpha}{\pi} + \frac{\alpha\,\pi}{4}\,\beta + \mathcal{O}(\beta^2)\,, \tag{11}$$

$$\delta F_1^{\rm IR} = -\frac{2}{3}\frac{\alpha}{\pi}\,\beta^2 + \mathcal{O}(\beta^4)\,. \tag{12}$$

The leading term of $F_1^{\rm fin}$ is proportional $\pi\alpha/\beta$ and exhibits the familiar Coulomb singularity. Also the constant term and the term linear in $\beta$ are infrared finite. The infrared singular part of $F_1$ is strongly suppressed at threshold $\propto \beta^2$, giving rise to a $\beta^3$ contribution to the rate. The correction to the Pauli form factor $\delta F_2$ is infrared finite and approaches a constant value at threshold:

$$\delta F_2 = -\frac{1}{2}\frac{\alpha}{\pi} + \mathcal{O}(\beta^2)\,. \tag{13}$$

Real radiation, in contrast, vanishes as $\beta^3$ in the threshold region, where two powers of $\beta$ result from the square of the dipole matrix element, and one power of $\beta$ comes from phase space. It exhibits the same logarithmic dependence on the infrared cutoff as the $F_1$ form factor and the same leading $\beta$ dependence as the infrared singular part of the virtual correction. As a consequence of the strong suppression $\propto \beta^3$ it can be neglected in the threshold region, together with the corresponding infrared divergent part of the form factor. The angular distribution and, similarly, the correction to the total cross section in the threshold region are therefore determined by the the infrared finite parts of the form factors. To order $\alpha$ one thus finds for the coefficient describing the angular dependent piece

$$A = A_{\rm Born}\left(1 + \frac{2}{2-\beta^2}\,\frac{\alpha}{\pi}\right) . \tag{14}$$

As we shall show there are interesting modifications of the anisotropy due to the running of the QCD coupling, and the dependence of the renormalization scale on $\sqrt{s}$ and $|\vec{q}|$ will be crucial.

The $\mathcal{O}(\alpha^2)$-QED corrections to the form factors, induced by light fermion loops, have been calculated analytically in [3]. In the threshold region one obtains

$$\begin{aligned} F_1 &= 1 + \frac{\alpha\,\pi}{4\,\beta}\left[1 + \left(\frac{\alpha}{\pi}\right)\sum_{i=1}^{n_f}\frac{1}{3}\left(\ell n\frac{s\,\beta^2}{m_i^2} - \frac{8}{3}\right)\right] \\ &\quad - \frac{3}{2}\,\frac{\alpha}{\pi}\left[1 + \left(\frac{\alpha}{\pi}\right)\sum_{i=1}^{n_f}\frac{1}{3}\left(\ell n\frac{s}{4\,m_i^2} - \frac{1}{2}\right)\right] , \end{aligned} \tag{15}$$

$$\begin{aligned} F_2 &= \frac{\alpha\,\pi}{4\,\beta}\left[\left(\frac{\alpha}{\pi}\right)\frac{n_f}{3}\right] \\ &\quad - \frac{1}{2}\,\frac{\alpha}{\pi}\left[1 + \left(\frac{\alpha}{\pi}\right)\sum_{i=1}^{n_f}\frac{1}{3}\left(\ell n\frac{s}{4\,m_i^2} - \frac{13}{6}\right)\right] . \end{aligned} \tag{16}$$

The calculation has been performed in the limit where the mass of the light virtual fermion $m_f$ is far smaller than $m$, a situation appropriate for the subsequent translation to QCD. The factor $n_f$ is introduced to allow for several light fermions and, in our case, to single out the fermion-induced terms. These formulae provide the first step on the way to a full two-loop calculation in order $\alpha^2$. As we shall see, the results require two conceptionally different scales in the argument of the running coupling, a scale of order $s$ from the hard virtual correction from transverse photons and a soft scale of order $s\,\beta^2$ from the Coulomb rescattering. Supplemented by the BLM prescription they even determine the dominant two-loop gluon-induced terms in QCD.

The linear combination appearing in the denominator of $\widetilde{A}$ in Eq. (6) is thus given by

$$\begin{aligned} F_1 + F_2 &= 1 + \frac{\alpha\,\pi}{4\,\beta}\left[1 + \left(\frac{\alpha}{\pi}\right)\sum_{i=1}^{n_f}\frac{1}{3}\left(\ell n\frac{s\,\beta^2}{m_i^2} - \frac{5}{3}\right)\right] \\ &\quad - 2\,\frac{\alpha}{\pi}\left[1 + \left(\frac{\alpha}{\pi}\right)\sum_{i=1}^{n_f}\frac{1}{3}\left(\ell n\frac{s}{4\,m_i^2} - \frac{11}{12}\right)\right]. \end{aligned} \tag{17}$$

The $n_f$ terms arise from the vacuum polarization insertions and thus can be resummed into the QED running coupling:

$$\alpha(Q^2) = \alpha\left[1 + \left(\frac{\alpha}{\pi}\right)\sum_{i=1}^{n_f}\frac{1}{3}\left(\ell n\frac{Q^2}{m_i^2} - \frac{5}{3}\right)\right]. \tag{18}$$

The constant 5/3 is the usual term in the Serber-Uehling vacuum polarization $\Pi(Q^2)$ at large $Q^2$. This corresponds to the usual QED scheme where $V(Q^2) = -4\pi\,\alpha(Q^2)/Q^2$ is the QED potential for the scattering of heavy test charges. One thus obtains

$$\begin{aligned} F_1 + F_2 &= 1 + \frac{\alpha(s\,\beta^2)\,\pi}{4\,\beta} - 2\,\frac{\alpha(s\,e^{3/4}/4)}{\pi} \\ &\cong \left(1 - 2\,\frac{\alpha(s\,e^{3/4}/4)}{\pi}\right)\left(1 + \frac{\alpha(s\,\beta^2)\,\pi}{4\,\beta}\right). \end{aligned} \tag{19}$$

Two distinctly different correction factors arise. The first originates from hard transverse photon exchange, with the scale set by the short distance process; the second from the instantaneous Coulomb potential. It is remarkable and non-trivial that the non-logarithmic terms in the $\pi\alpha/\beta$ corrections are absorbed if the relative momentum is adopted as the scale for the coupling. Up to two loops the running coupling governing the Coulomb singularity is thus identical to the running coupling in the potential. This will provide an important guide for the application of these results to QCD.

The proper resummation of the $1/\beta$ terms based on Sommerfeld's rescattering formula then leads to

$$|F_1 + F_2|^2 = \left(1 - 4\,\frac{\alpha(m^2\,e^{3/4})}{\pi}\right)\frac{x}{1 - e^{-x}} \tag{20}$$

with

$$x = \frac{\alpha(4\,m^2\,\beta^2)\,\pi}{\beta}\,. \tag{21}$$

In a similar way one finds for the relevant combination in the numerator of (6)

$$|F_1|^2 - |F_2|^2 \cong \left(1 - 3\,\frac{\alpha(m^2\,e^{7/6})}{\pi}\right)\,\frac{x'}{1-e^{-x'}} \tag{22}$$

with

$$x' = \frac{\alpha(4\,m^2\,\beta^2/e)\,\pi}{\beta}\,. \tag{23}$$

The $|F_2|^2$ term in the numerator can actually be ignored in the present approximation. The scales of the effective coupling differ between the numerator and denominator of (6): In particular in the factor arising from Coulomb exchange the scale is significantly smaller in the numerator than in the denominator. This behavior is consistent with qualitative considerations based on the relative distances relevant for $S$- versus $P$-waves in the Coulomb part. In the factor arising from hard photon exchange the scales are quite comparable, with a slightly larger value in the numerator.

One thus arrives at the prediction in the context of QED for the anisotropy which involves four scales:

$$A = \frac{\widetilde{A}}{1-\widetilde{A}}\,, \qquad \widetilde{A} = \frac{\beta^2}{2}\,\frac{\left(1 - 3\,\frac{\alpha(m^2 e^{7/6})}{\pi}\right)}{\left(1 - 4\,\frac{\alpha(m^2 e^{3/4})}{\pi}\right)}\,\frac{1-e^{-x}}{1-e^{-x'}}\,\frac{\alpha(4\,m^2\,\beta^2/e)}{\alpha(4\,m^2\,\beta^2)}\,. \tag{24}$$

To display the effects more clearly, the ratio of the anisotropy to the Born prediction $A/A_{\rm Born}$ is shown in Fig. 1 for the case of $\tau$ pair production. The dashed curve gives the prediction for constant $\alpha_{\rm QED}$; the solid curve shows the effect of the lepton vacuum polarization $\Pi(Q^2)$ in the QED running coupling. The vacuum polarization affects the anisotropy for small $\beta$ because two different scales appear in the $S$- and $P$-wave Coulomb rescattering corrections. Away from threshold $A$ essentially measures the anomalous magnetic moment.

## 3. From QED to QCD

The QED coupling $\alpha(Q^2)$ translates into the the QCD coupling $\alpha_{\rm V}(Q^2)$, defined as the effective

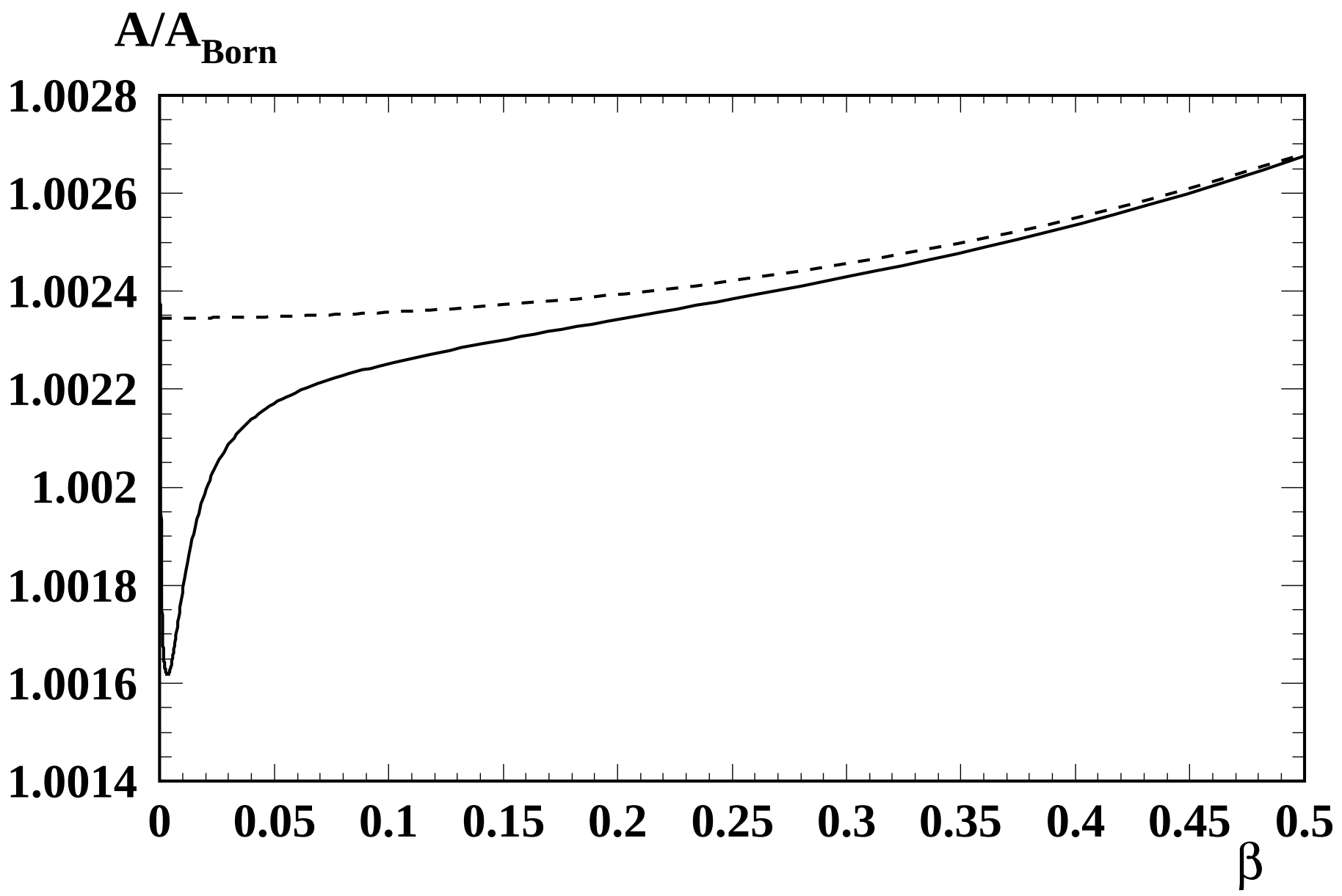

Figure 1: Ratio between the anisotropy $A$ and the Born prediction $A_{\rm Born}$ as function of $\beta$ for the process $e^+e^- \to \tau^+\tau^-$. Dashed curve: constant $\alpha$; solid curve: including the running of $\alpha$.

charge in the potential

$$V(Q^2) = -\,\frac{4}{3}\,\frac{4\,\pi\,\alpha_{\rm V}(Q^2)}{Q^2} \tag{25}$$

for the scattering of two heavy quarks in a color-singlet state. In the BLM procedure all terms arising from the non-zero beta-function are resummed into $\alpha_{\rm V}(Q^2)$. For example, all $n_f$-dependent coefficients vanish in the $\pi\alpha/\beta$ terms if the scale of the relative momentum is adopted. This is, in fact, a result expected on general grounds: threshold physics is governed by the nonrelativistic instantaneous potential. Below threshold, the potential leads to bound states, above threshold it affects the cross section through final state interactions. It is, therefore, natural to take for the QCD case the coupling governing the QCD potential at the momentum scale involved in the rescattering. To relate $\alpha_{\rm V}$ to $\alpha_{\overline{MS}}$, we use

$$\alpha_{\overline{MS}}(Q^2) = \alpha_{\rm V}(e^{+5/3}\,Q^2)\,\left[1 + 2\,\frac{\alpha_{\rm V}}{\pi} + \mathcal{O}(\alpha_{\rm V}^2)\right]\,. \tag{26}$$

In a similar way, BLM scale-fixing is adopted for the correction from hard gluon exchange. In the radiative correction, there still remain $\mathcal{O}(\alpha_s^2)$ terms, identical to the radiative corrections for the theory with a fixed coupling constant. With the same scheme convention for the coupling as above, one arrives at

$$\widetilde{A} = \frac{\beta^2}{2} \frac{\left(1 - 4\, \frac{\alpha_{\rm V}(m^2 e^{7/6})}{\pi}\right)}{\left(1 - \frac{16}{3}\, \frac{\alpha_{\rm V}(m^2 e^{3/4})}{\pi}\right)} \, \frac{1 - e^{-x_s}}{1 - e^{-x_s'}} \, \frac{\alpha_{\rm V}(4\, m^2 \beta^2 / e)}{\alpha_{\rm V}(4\, m^2 \beta^2)} \tag{27}$$

where

$$x_s = \frac{4\,\pi}{3} \frac{\alpha_{\rm V}(4\, m^2 \beta^2)}{\beta}, \qquad x_s' = \frac{4\,\pi}{3} \frac{\alpha_{\rm V}(4\, m^2 \beta^2 / e)}{\beta}. \tag{28}$$

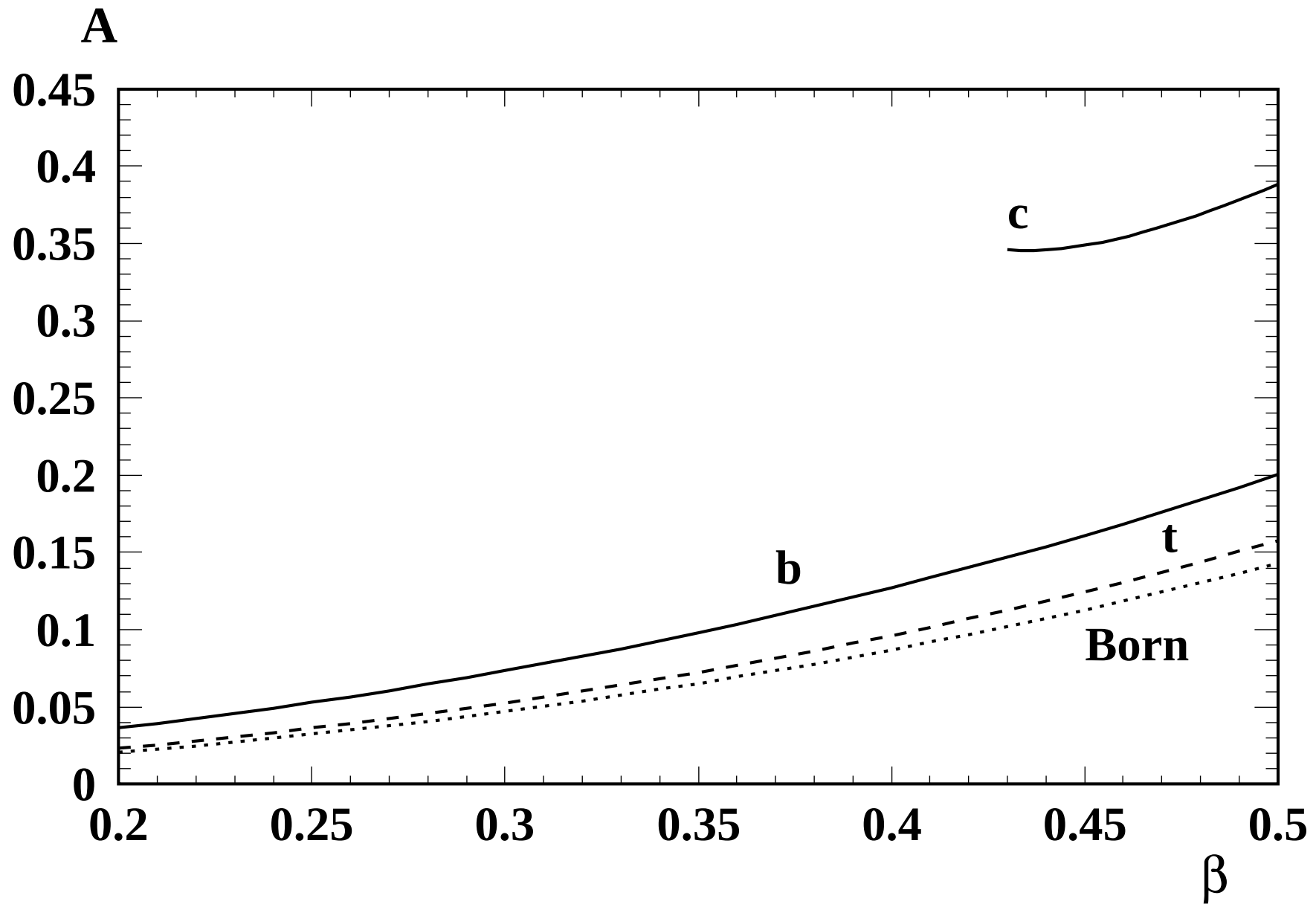

Figure 2: Anisotropy for charmed, bottom and top quark production as a function of $\beta$. Also shown is the Born prediction. We have assumed the effective quark masses $m_c = 1.7$ GeV, $m_b = 5$ GeV and $m_t = 175$ GeV.

The anisotropy $A$ is plotted in Fig. 2 versus the velocity $\beta$ in the range $0.2 < \beta < 0.5$ for charmed, bottom, and top quarks. For comparison, the tree level prediction is also shown. For charmed quarks, only $\beta$ values above 0.4 are admitted in order to allow for the simultaneous production of $D\overline{D}$ and $D^*\overline{D}$. The charm prediction is particularly sensitive to the QCD parameters, since very low scales are accessible. Measurements of the anisotropy for $e^+e^- \to c\overline{c}$ thus have the potential of determining $\alpha_{\rm V}$ in the regime where perturbation theory begins to fail.

The curves are based on an input value $\alpha_{\overline{MS}}^{(n_f=5)}(M_Z^2) = 0.115$. We use the two-loop beta-function to evolve $\alpha_{\overline{MS}}$ to lower momenta and then used Eq. (26) to calculate $\alpha_V(Q^2)$. To investigate the sensitivity of the predictions for bottom quarks, the input value for $\alpha_{\overline{MS}}(M_Z^2)$ has been varied by $\pm 0.008$ from the central value of 0.115. As demonstrated in Fig. 3a the variation of the anisotropy parameter amounts to about 10%, and could therefore be accessible experimentally. The charm predictions (see Fig. 3b) are even more sensitive.

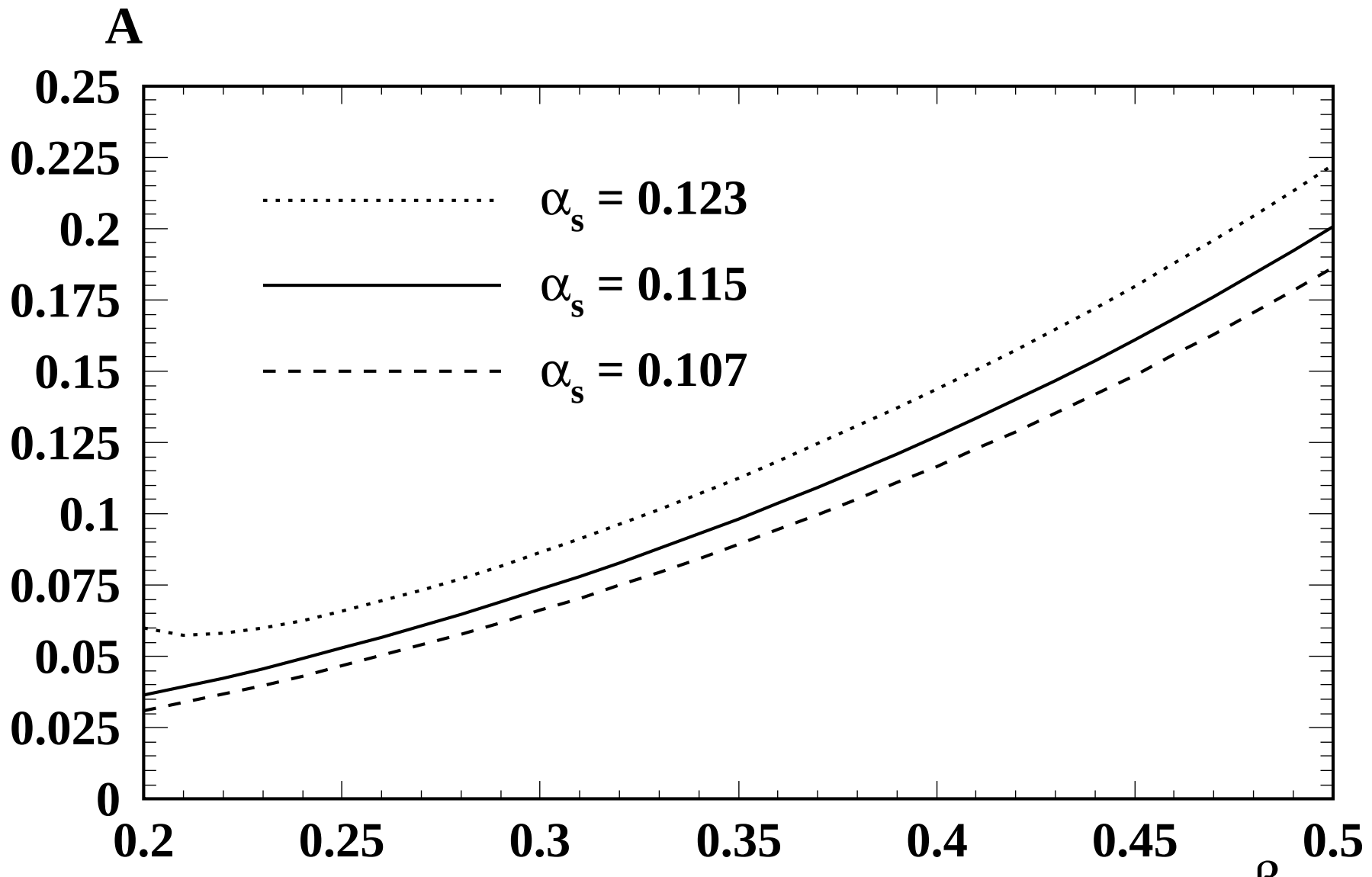

(a)

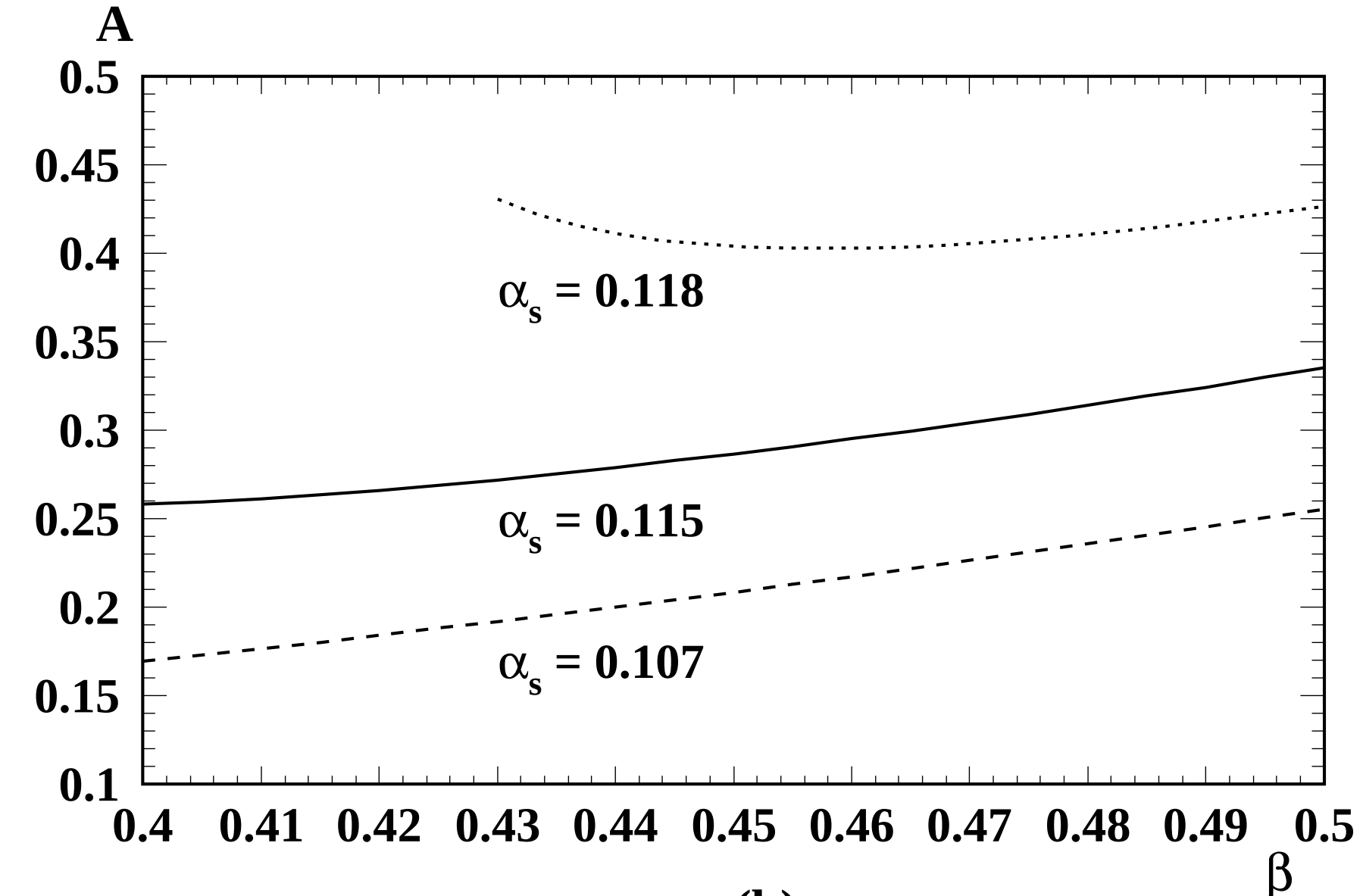

(b)

Figure 3: Sensitivity of the anisotropy $A$ for (a) $e^+e^- \to b\,\overline{b}$ and (b) $e^+e^- \to c\,\overline{c}$ to changes in $\alpha_{\overline{MS}}(M_Z^2)$.

## 4. Conclusions

An important consequence of heavy quark kinematics is that the production angle of a heavy hadron follows the direction of the parent heavy quark. This applies not only at Born approximation, but also after QCD corrections have been applied. In this paper we have shown that the anisotropy $A(\beta^2)$ in the cm angular distribution $d\sigma(e^+e^- \to Q\overline{Q})/d\Omega \propto 1 + A\cos^2\theta$ of heavy quarks produced near threshold is sensitive to the QCD coupling $\alpha_V(Q^2)$ at specific scales determined by the quark relative momentum $p_{\rm cm} = \sqrt{s}\beta$. The coupling $\alpha_V(Q^2)$ is the physical effective charge defined through heavy quark scattering. The predictions provide a connection between observables and thus are independent of theoretical conventions.

An important feature of our analysis is the use of BLM scale-fixing, in which all higher-order corrections associated with the beta-function are resummed into the scale of the coupling. The resulting scale for $\alpha_V(Q^2)$ corresponds to the mean gluon virtuality. In the case of the soft rescattering corrections to the $S$-wave, the BLM scale is $s\beta^2 = p_{\rm cm}^2$. One thus has sensitivity to the running coupling over a range of momentum transfers within the same experiment. The anisotropy measurement thus can provide a check on other determinations of $\alpha_V(Q^2)$, e.g. from heavy quark lattice gauge theory, or from the conversion of $\alpha_{\overline{MS}}$ determinations to $\alpha_V$ as given in Eq. (26).

Our analysis also shows that the running coupling appears within the cross section with several different scales. This is particularly apparent at low $\beta$ where the physical origin of the $\mathcal{O}(\alpha_s)$ corrections can be traced to gluons with different polarization and virtuality.

In principle, the anisotropy of $\tau$ pairs produced in $e^+e^- \to \tau^+\tau^-$ could be used to measure the Pauli form factor $F_2(s)$ near threshold $s \gtrsim 4m_\tau^2$. A highly precise measurement of the anisotropy thus could provide a measurement of a fundamental parameter of the $\tau$ lepton, its timelike anomalous magnetic moment.

**Acknowledgments**

One of the Authors (JHK) would like to thank the SLAC Theory Group for hospitality and the Volkswagen-Stiftung Grant I/70452 for generous support.